# Framework of compressive sensing and data compression for 4D-STEM

Hsu-Chih Ni[a,b], Renliang Yuan[c], Jiong Zhang[c], Jian-Min Zuo[a,b,*]

a Department of Materials Science and Engineering, University of Illinois at Urbana-Champaign, Urbana, IL 61801, USA

b Materials Research Laboratory, University of Illinois at Urbana-Champaign, Urbana, IL 61801, USA

c Intel Corporation, Corporate Quality Network, Hillsboro, OR 97124, USA

*Corresponding author email, jianzuo@illinois.edu

## Abstract

Four-dimensional Scanning Transmission Electron Microscopy (4D-STEM) is a powerful technique for high-resolution and high-precision materials characterization at multiple length scales, including the characterization of beam-sensitive materials. However, the field of view of 4D-STEM is relatively small, which in absence of live processing is limited by the data size required for storage. Furthermore, the rectilinear scan approach currently employed in 4D-STEM places a resolution- and signal-dependent dose limit for the study of beam sensitive materials. Improving 4D-STEM data and dose efficiency, by keeping the data size manageable while limiting the amount of electron dose, is thus critical for broader applications. Here we develop a general method for reconstructing 4D-STEM data with subsampling in both real and reciprocal spaces at high fidelity. The approach is first tested on the subsampled datasets created from a full 4D-STEM dataset, and then demonstrated experimentally using random scan in real-space. The same reconstruction algorithm can also be used for compression of 4D-STEM datasets, leading to a large reduction (100 times or more) in data size, while retaining the fine features of 4D-STEM imaging, for crystalline samples.

## 1. Introduction

Scanning transmission electron microscopy (STEM) is a powerful tool for materials characterization. By scanning a focused electron probe across a sample region and collecting scattered electrons using a detector, a real-space image is formed in various forms of contrast, from atomic resolution Z-contrast to annular bright-field imaging of light atoms to differential phase contrast[1,2]. However, only a part of the scattered electrons is collected, and the electron momentum isn't resolved, or only partially resolved. Most of diffraction information about the material is thus lost in STEM imaging. The emerging four-dimensional (4D) STEM or 4D-STEM collects electron diffraction patterns (DPs) at each probe position using a pixelated detector. This technique combines the advantages of electron diffraction with STEM imaging, which is becoming more and more popular due to the advent of new detector technologies[3–6]. Studies have shown that a number of materials structural properties can be mapped by 4D-STEM, including but not limited to strain mapping[7–9], orientation mapping[10–12], and atomic arrangements using atomic resolution ptychography[13,14].

However, current 4D-STEM analysis is constrained by several factors. The first is the amount of beam dwell time to record a DP, which is milliseconds or more with the current generation of direct electron detectors[3,4,6]. The dwell time could be even longer using cameras with a large pixel format and a limited data readout rate. Coupled with the minimum dose required to record a DP with the required signal and noise ratios place a lower limit on the electron dose rate. Typically, the amount of electron dose on the sample in a 4D-STEM experiment can be orders larger compared with conventional STEM. Second, for a fixed pixel resolution, the size of 4D-STEM datasets is proportional to the scan size, which determines the field of view (FOV), and thus the areas that can be analyzed by 4D-STEM is practically limited by the data size, in addition to long acquisition times. These limits must be overcome to improve the throughput of 4D-STEM analysis and to extend 4D-STEM applications to a broad range of materials, especially, to samples susceptible to beam damage.

The current implementation of 4D-STEM samples the real and reciprocal space uniformly using the STEM raster scan and a pixelated area detector. The amount of sampling follows the Nyquist sampling theorem at a rate that is twice the Nyquist frequency or higher. This sampling strategy ignores spatial correlations at different sample locations and spectral correlations among DPs. It creates massive amounts of data that are only to be used partially for image formation. This process is thus extremely inefficient and wasteful in the use of microscopy and computational resources and the electron dose.

Compressive sensing (CS) [15–26] offers an alternative sampling strategy against the traditional Nyquist sampling. This strategy has attracted large interest in the electron microscopy community for overcoming electron dose limitations in the STEM spectroscopy analysis lately[16,27–29]. The general idea is that signals obtained from samples are often sparse and its representation as a high-dimensional vector could be described by a small number of basis signal vectors (or endmembers).  In such cases, an incoherent sample with a sampling rate much lower than the Nyquist sampling limit could be sufficient for recovering the original signal[22,30]. Applications have demonstrated that CS could greatly reduce the amount of the electron dose and acquisition time for STEM imaging and spectroscopy mapping. For example, Andrew et al.[18] developed a compressive sensing method based on the beta-process factor-analysis and applied it on the beam sensitive ZSM-5 zeolite sample. Their results showed that with only 10% of sampling, most of the information regarding the atomic columns and varying contrast in the ZSM-5 zeolite was retained in the reconstructed STEM images. Compressive sensing has also been explored for STEM tomography to reduce the number of sampling points and thus lowering the tilt-series acquisition time [26] A strategy for acquiring and restoring the sub-sampled STEM tomography datasets with large tilt steps and partially scanned STEM images was demonstrated by Saghi et al. [26]. For electron spectroscopy, Monier et al.[31] reconstructed randomly sampled EELS maps by regularizing spatial smoothness and enforcing spectral sparsity by hard-thresholded principal component analysis (PCA). An EELS spectra map of biological tissue was obtained with the method.

Another bonus of CS is dataset compression, which is naturally made. With fast detectors and opportunities for acquiring datasets from large FOVs, a 4D-STEM experiment entails large volumes of data. The storage of these large datasets is problematic, especially when multiple analyses are made. This emerging data issue has been discussed under the context of direct electron detectors and cryogenic electron microscopy [32], but has yet to be addressed regarding 4D-STEM. In CS, data can be decomposed into a small set of base vectors and their coefficients. The sizes of these two matrices can be significantly smaller than the original dataset, and thus greatly reduce the amount of stored data. To find the base vectors and their coefficients, a decompression algorithm is needed. This problem is the same as reconstruction used in hyperspectral compressive imaging, for which different algorithms have been proposed[33–37].

In this paper, we introduce a dual-space (real and reciprocal spaces) CS (DSCS) scheme for 4D-STEM and demonstrate its use in compressive sensing for data collection and data compression of 4D-STEM datasets. The data collected by 4D-STEM is similar to hyperspectral images (HSI) in remote sensing, in the sense that both form a stack of images using the spectral signals of specific wavelength as in HSI and of electron momentum in 4D-STEM. The spatial-spectral compressed reconstruction based on the spectral unmixing (SSCR_SU) algorithm proposed by Wang et al. [33] originally for remote sensing is adapted and developed for 4D-STEM. The DSCS takes the sampled STEM images and DPs and uses them as input to reconstruct the full 4D-STEM dataset. We explore two sampling schemes. Real-space random scan based on Bernoulli sampling is used to obtain a set of DPs in both schemes. In the spectral (diffraction) space, one scheme uses images formed using segmented STEM detectors, and the other scheme uses random detector pixels for imaging. The DPs and images are collected separately. The use of segmented STEM detectors simplifies the DSCS implementation since it only requires a STEM capable of random scan. Once the DSCS datasets are acquired, endmembers are extracted from the collected DPs, while the abundance of each endmember is estimated from the recorded STEM images. We then combine the endmembers and their abundance to reconstruct the full 4D-STEM data. We compare the reconstruction results using the structural similarity index (SSIM) and peak signal to noise ratio (PSNR). The effectiveness of the DSCS and SSCR_SU algorithms in the endmember and abundance estimations, and the accuracy of reconstructed 4D-STEM datasets, are demonstrated using strain mapping as an application test.

## 2. DSCS and Reconstruction

A critical assumption for CS is that the data is sparse. For a 4D-STEM experiment, sparsity means that each DP can be separated into several "endmembers", each endmember has the same dimension as the original DPs. Importantly, the number of endmembers is much smaller than the number of DP pixels. That is, DPs can be described with only a handful of parameters and a set of endmembers. Also, we assume that DPs are spatially correlated so that these endmembers are shared between DPs in the FOV. Another assumption is that DPs can be represented as linear combinations of endmembers. The coefficients of linear combinations can be mapped in real-space and hereinafter referred as the abundance matrix $S$. The 4D-STEM dataset can be considered as the matrix multiplication of

$$X = SA \tag{1}$$

Where $X \in R^{N \times L}$ is the full 4D-STEM dataset, $A \in R^{p \times L}$ is the endmember matrix and $S \in R^{N \times p}$ is the abundance matrix. $N$ is the number of pixels in the FOV, $L$ is the number of pixels per DP and $p$ is the number of endmembers.

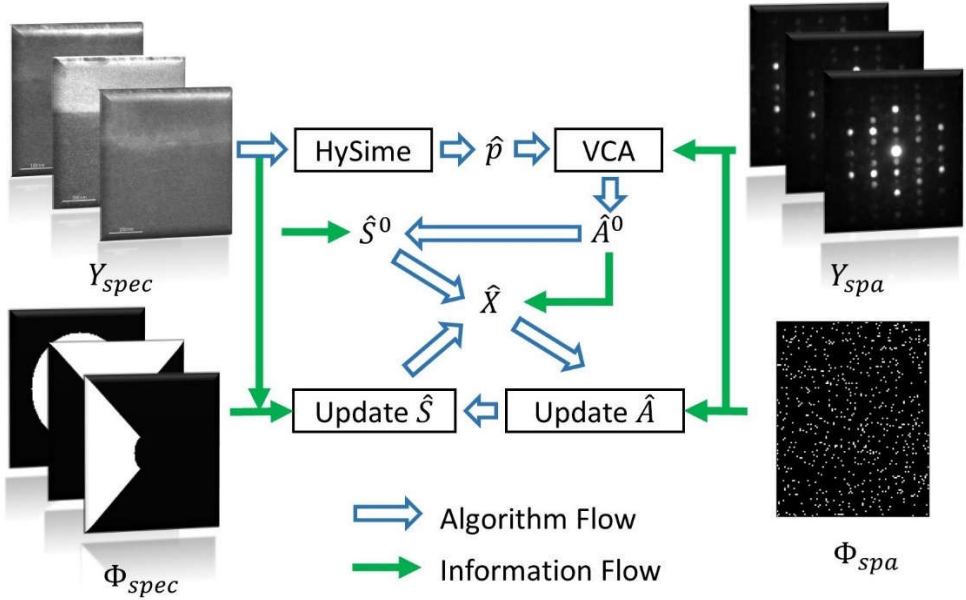

*Figure 1. 4D-STEM compressive sensing and reconstruction algorithm based on SSCR_SU. The number of endmembers $\hat{p}$ is first estimated from $Y_{spec}$ (the STEM images) stack using HySime. Then VCA is used with $Y_{spa}$ (DPs) to estimate the endmembers matrix $\hat{A}^0$. Abundance matrix $\hat{S}^0$ is then estimated. The multiplication of $\hat{S}^0$ and $\hat{A}^0$ gives $\hat{X}$, Which is the estimated full 4D-STEM dataset. $\hat{X}$ is subsequently optimized by updating $\hat{A}$ and $\hat{S}$ sequentially.*

We found the SSCR_SU algorithm developed by Wang et al. [33] can be adapted for 4D-STEM. SSCR_SU combines *spectral sampling* with *spatial sampling* for CS. Spectral sampling is achieved by STEM imaging where diffraction intensity serves as the spectral signal. The SSCR_SU reconstruction algorithm (Fig. 1) retrieves information from spatial sampling (real space) and spectral sampling (diffraction space). The reconstruction takes four inputs: 1) DPs of each sampled position ($Y_{spa}$), 2) a stack of STEM images acquired with different STEM detector configurations and different camera lengths ($Y_{spec}$), 3) the sampling positions ($\Phi_{spa} \in R^{n_{spa} \times N}$), and the 4) the regions on the DP which are spectrally sampled ($\Phi_{spec} \in R^{n_{spec} \times L}$). Here, $n_{spa}$ and

$n_{spec}$ are the number of samples in real space and spectral (diffraction) space. The relationships between these inputs are

$$Y_{spa} = \Phi_{spa} X \tag{2}$$

and

$$Y_{spec} = X \Phi_{spec}^{T} \tag{3}$$

Where X is the full 4D-STEM dataset. $\Phi_{spa}$ is an identity matrix of size N for a raster scan when all pixels are probed. In DSCS, we randomly select $n_{spa}$ sampling points with $n_{spa} \ll N$, through Bernoulli sampling, which is equivalent to randomly select $n_{spa}$ rows from the identity matrix of size N to form $\Phi_{spa}$. The random sampling in principle can be employed for spectral sampling as well, which works in the case of 4D-STEM data compression. The experimental implementation of random spectral sampling, however, requires the design of detectors capable of random pixel readout, which can be developed in future. DSCS can also be implemented using the available STEM detectors, for example, using a BF and 4 segmented STEM detectors (Fig. 1).

The reconstruction algorithm first estimates the number of endmembers ($\hat{p}$) in the stack of STEM images using the HySime algorithm[38], as shown in Fig. 1. These images contain information about the real-space distribution of the sampled spectral signals. It can be considered as a spectrally compressed 4D-STEM dataset, and thus it serves as a surrogate for the full 4D-STEM dataset. After obtaining $\hat{p}$, we perform vertex component analysis (VCA) [39] to estimate the initial endmember matrix $\hat{A}^0$ from the sampled DPs. Next, a first estimation of the abundance matrix $\hat{S}^0$ is obtained by multiplying the stack of STEM images with the pseudo inverse of the spectrally sampled endmember matrix.

$$\hat{S}^0 = Y_{spec}(\hat{A}^0 \Phi_{spec}^T)^T \left[(\hat{A}^0 \Phi_{spec}^T)(\hat{A}^0 \Phi_{spec}^T)^T\right]^{-1}, \tag{4}$$

with the initial estimation of the full 4D-STEM dataset given by

$$\hat{X}^0 = \hat{S}^0 \hat{A}^0. \tag{5}$$

This estimation is further improved by updating $S^k$ and $A^k$ through the optimization of

$$A^k = \underset{A}{\operatorname{argmin}} \|\hat{X}^{k-1} - S^{k-1} A\|_2^2 + \lambda_1 \|Y_{spa} - \Phi_{spa} S^{k-1} A\|_2^2 \tag{6}$$

and

$$S^k = \underset{S}{\operatorname{argmin}} \|\hat{X}^{k-1} - S A^k\|_2^2 + \lambda_2 \|Y_{spec} - S A^k \Phi_{spec}^T\|_2^2 \tag{7}$$

where

$$\hat{X}^k = S^k A^k, \tag{8}$$

and $k$ denotes the number of iterations and $\|\cdot\|_2$ denotes the Frobenius norm of the matrix. $S$ and $A$ are optimized alternatively in an iterative loop. $\lambda_1$ and $\lambda_2$ are weighting parameters for the spatial and spectral constraints. The above step is called "joint optimization". The optimization ends when a certain value or the relative change is smaller than a preset criterion. The relative change is defined as

$$relative\ change = \frac{\|\hat{X}^k - \hat{X}^{k-1}\|_2}{\|\hat{X}^k\|_2}. \tag{9}$$

For the reconstruction performance evaluation, we use Peak signal to noise ratio (PSNR)[40] :

$$PSNR(I, \hat{I}) = -20 \log_{10} \left( \frac{\max(I)}{\frac{1}{N}\|I - \hat{I}\|_2} \right). \tag{10}$$

Higher PSNR value suggests the mean squared error is smaller, relative to the max intensity in the true image. Another performance index is the structural similarity index, which is defined by [41]

$$SSIM(I, \hat{I}) = \frac{(2\mu_I \mu_{\hat{I}} + C_1)(2\sigma_{I\hat{I}} + C_2)}{(\mu_I^2 + \mu_{\hat{I}}^2 + C_1)(\sigma_I^2 + \sigma_{\hat{I}}^2 + C_2)}, \tag{11}$$

where $\mu_i$ is the average of $i$, $\sigma_i$ is the variance of $i$, $C_1 = (0.01L)^2$ and $C_2 = (0.03L)^2$, $L$ is the dynamic range of the pixel value. The range of SSIM is 0 to 1, and higher SSIM means higher similarity between two images. Detail discussion of the metrics can be found in [40,41]

The DSCS algorithm described here is implemented in Python. All reconstructions and testing are done on a personal computer with an Intel i7-12700 CPU and 32 GB of RAM. Two different quantitative indexes are employed to evaluate the quality of reconstruction.

## 3. Results and Discussions

### 3.1 Creation of sub-sampled 4D-STEM dataset

We use a 4D-STEM dataset collected using raster scan from a nanowire transistor to test our DSCS framework. The TEM sample was lift-out by focused-ion beam (FIB) from a p-MOSFET device manufactured by Interuniversity Microelectronics Centre (IMEC). The sample is oriented near the [110] zone axis with a thickness around 40 to 50 nm. The selected FOV contains a nanowireFET, including two nanowire channel, epitaxy SiGe stressor and Si substrate body. The nanowire channel is covered with gate materials. (Fig. 2).

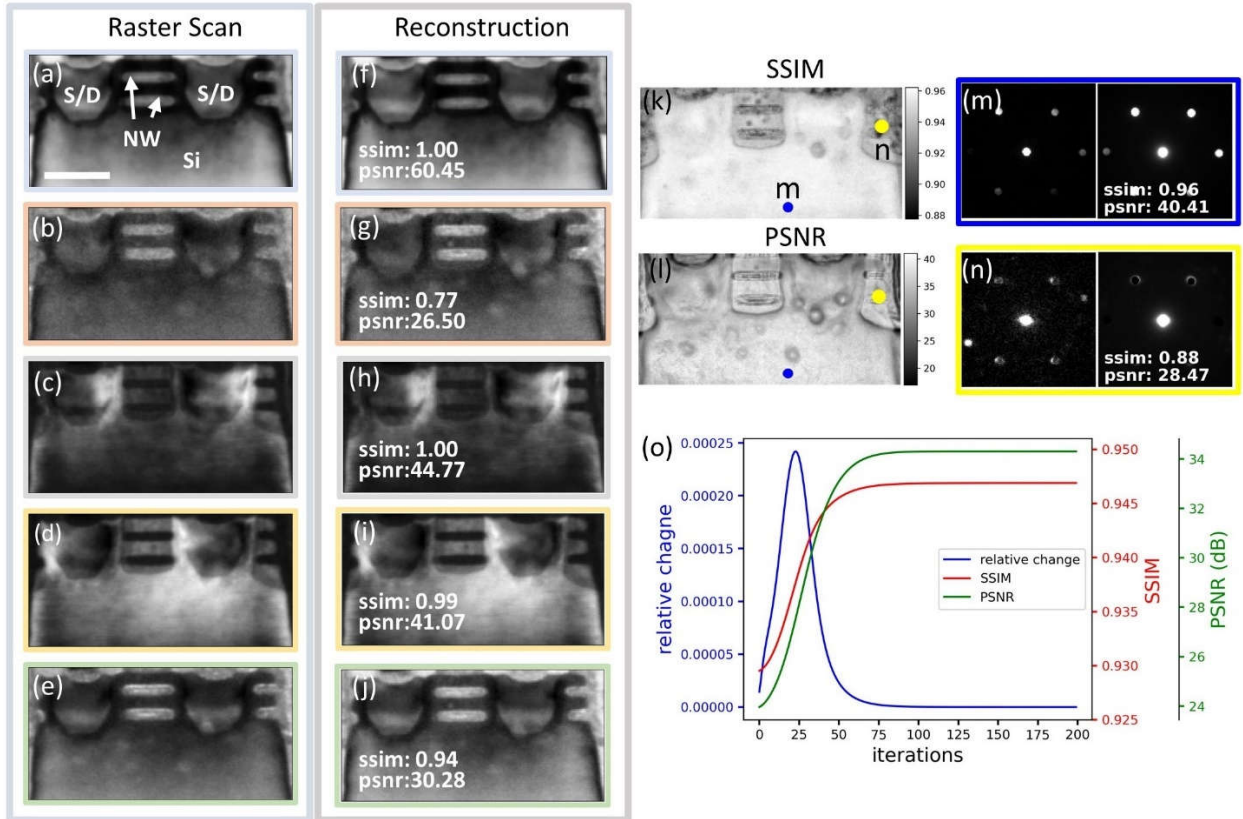

*Figure 2 STEM images created with virtual detectors. (a-e) virtual DF images created from the original 4D-STEM dataset. The positions of source/drain (S/D), nanowire channel (NW), and Si subfin (Si) is labeled in (a). (f-j) virtual DF images created from the reconstructed 4D-STEM dataset. The size of the detectors used for (a-j) corresponds to camera length of 4300 mm. (k, l) SSIM and PSNR map between the original dataset and the reconstructed dataset. The value of each points represents the similarity metrics value between the original and reconstructed DP. (m,n) DPs extracted from original and reconstructed datasets. The spatial positions of these DPs are marked as yellow and red dot in (k,l), respectively. (o) Change of average similarity metrics with respect to number of iterations of joint optimization.*

The 4D-STEM dataset tested here was acquired with EMPAD[3] on a Talos F200X S/TEM (Thermo-Fisher Scientific, USA, operated at 200kV). We operated the microscope under the microprobe mode to form a probe of 0.6 mrad semi-convergence angle and diameter 2.2 nm in full-width half-maximum. The exposure time was 1 *ms* for each DP. The DPs were taken near the [110] zone axis, and the camera length was adjusted so only {002} and {1$\bar{1}$1} diffraction disks were included in the recorded DPs.

We synthesized a subset of the full 4D-STEM dataset ($X_{syn}$) to test the DSCS reconstruction algorithm. The full 4D-STEM dataset has 200 by 100 pixels in the FOV, and each DP contains 124 by 124 pixels. The $X_{syn}$ was sampled with a spatial sample rate of 5%. Virtual bright-field (BF) and dark-field (DF) images of the dataset were created by mimicking the standard STEM detectors with one for BF and four segmented DF detectors. Each virtual detector is scaled to three different sizes, corresponding to the camera length of 2850, 3400 and 4300 mm, respectively. With these detector and camera settings, different parts of the DP are sampled, forming 15 virtual STEM images. Examples of virtual STEM images formed with different detector configurations are shown in Fig 2a-e. The diffraction contrast carried by virtual STEM images is encoded into the reconstruction dataset.

### 3.2 Reconstruction of the sub-sampled 4D-STEM dataset

The SSCR_RU reconstruction was performed using the synthesized dataset $X_{syn}$, producing a reconstructed 4D-STEM dataset $\hat{X}_{syn}$. Joint optimization was stopped when the relative change of Eq. 9 reached 1e-7. The virtual STEM images formed using the reconstructed dataset as shown in Fig. 2(f)-(j) are visually identical to the original virtual STEM images [(Fig. 2(a)-(e)]. The SSIMs of the two sets of images are all close to one. Figure 2k and l map out the SSIM and PSNR value for each DP in $\hat{X}_{syn}$, when compared with their counterpart in $X_{syn}$. The quality of the reconstructed DP is demonstrated using two pairs of DPs from $\hat{X}_{syn}$ and $X_{syn}$, respectively [(Fig. 2(m), (n)]. The DPs taken away from the interface show very high similarity [Fig. 2(m)], with the intensity variations among the diffraction disks reproduced in high fidelity. The reconstructed DPs also appear smoother because of the denoising effect during the process of endmember extraction. In comparison, the reconstructed DP closer to the interface of Si and $SiO_2$ [marked with a red cross in Fig. 2 (k) and (l)], have obvious artifacts. We also noticed dark contrast around the interfaces between crystalline Si and amorphous $SiO_2$ and the black circular contrast in the crystalline Si region. Those contrast are almost invisible in the virtual STEM images [(Fig. 2(a)-(e)], but very evident in the metrics maps. These circular regions are where electron beam was temporarily parked and were slightly contaminated. This introduces diffuse scattering in DPs, which aren't completely captured by the reconstruction algorithm.

We note that stopping criterion for joint optimization can influence the performance of the reconstruction algorithm. This dependence is measured using the relative change (Eq. 9) average SSIM and average PSNR for each iteration [Fig. 2(o)]. The reconstruction converges monotonically. The average SSIM and PSNR also increase, and they converge at ~80 iterations. The relative change negatively correlated with average SSIM and PSNR, indicating that relative change is a reasonable index for monitoring the optimization process. The criterion for stopping can be established as the point of convergence of relative change. Since the process of joint optimization is deterministic, the point of convergence can be determined by examining the relative change curve of a longer run. Here, the stopping criterion was set at relative change is equal to 1e-7, where both SSIM and PSNR are converged.

### 3.3 More than 1000-fold reduction in data size

The abundance matrix and endmember matrix are saved after reconstruction. For the 4D-STEM data described in Section 3.1 with 200x100 pixels in real space and DPs of 124x124 pixels, the dimension of abundance matrix and endmember matrix are 20000 by 6 and 6 by 15376, with 6 endmembers, respectively. Comparing with the full dataset, which is a matrix of 20000 by 15376, there is a 1400-fold reduction in the volume of data. This data reduction is very attractive as an efficient means for storing and transferring 4D-STEM dataset.

### 3.4 Influence of the spatial sampling rate

To examine the sampling rate effect, we created additional 10 sub-sampled datasets with increasing sampling rates. Fig. 3 plots the averaged SSIM and PSNR between the reconstructed and original DPs in the 4D-STEM dataset. There is a large improvement in reconstruction at the 5 percent sampling rate. This minimum sampling rate is likely to be dependent on the complexity of the sample. For the nanowire device sample tested in this paper, the DP stacks is mainly composed of Si (110) zone axis pattern, so they are relatively uniform despite the presence of strain. For samples with multiple grain or phases, higher spatial sampling rate may be required.

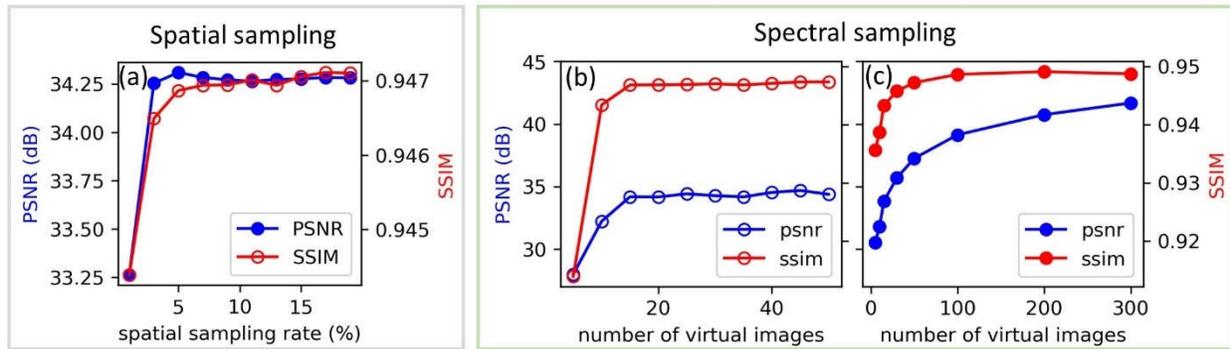

*Figure 3. Effects of spatial and spectral sampling on reconstruction performance. (a) influence of spatial sampling rate. Using (b) virtual STEM detector and (c) 10% randomly selected pixels for sampling in diffraction space and the influence of number of virtual images*

### 3.5 Influence of spectral sampling

A benchmark for spectral sampling configurations is shown in Fig. 3 (b). The virtual images, up to 50 of them, were created using different STEM detector configurations and the camera lengths of ranging from 2850 mm to 4300 mm. The improvements in the average SSIM and PSNR slow after more than 15 virtual images.

Alternatively, a very large number of virtual images can be created if random detector pixels are selected for STEM imaging. Each detector pixel creates a unique STEM image. We have also tested this spectral sampling scheme using our 4D-STEM test dataset. The virtual detectors are made of 10 percent of the total pixels in the DP, which are randomly selected. Fig. 3(c) shows that after having 30 or more detector configurations, random sampling outperforms STEM spectral sampling.

To further explore the effect of spectral sampling, we mapped the strain along the $(2\bar{2}0)$ direction from the original and the reconstructed 4D-STEM datasets and compared the two spectral sampling methods with 300 virtual images in both cases. The strain analysis was done with the procedure described by Yuan et al. [9]. Fig. 4 shows the results. Since strain analysis using 4D-STEM relies on accurately measuring the distances between diffraction disks, it provides a sensitive test of the DP quality.

Figure 4 shows while the strain maps obtained from the reconstructions using two different spectral sampling methods are similar and comparable to the strain map obtained from original 4D-STEM dataset, however, it is evident that random spectral sampling gives better results. By comparing the line profile of the strain map shown in Fig. 4(b), (d), and (f), we reached the same conclusion that random spectral sampling can reproduce the original result minus the difference that is comparable to noise fluctuation in original result.

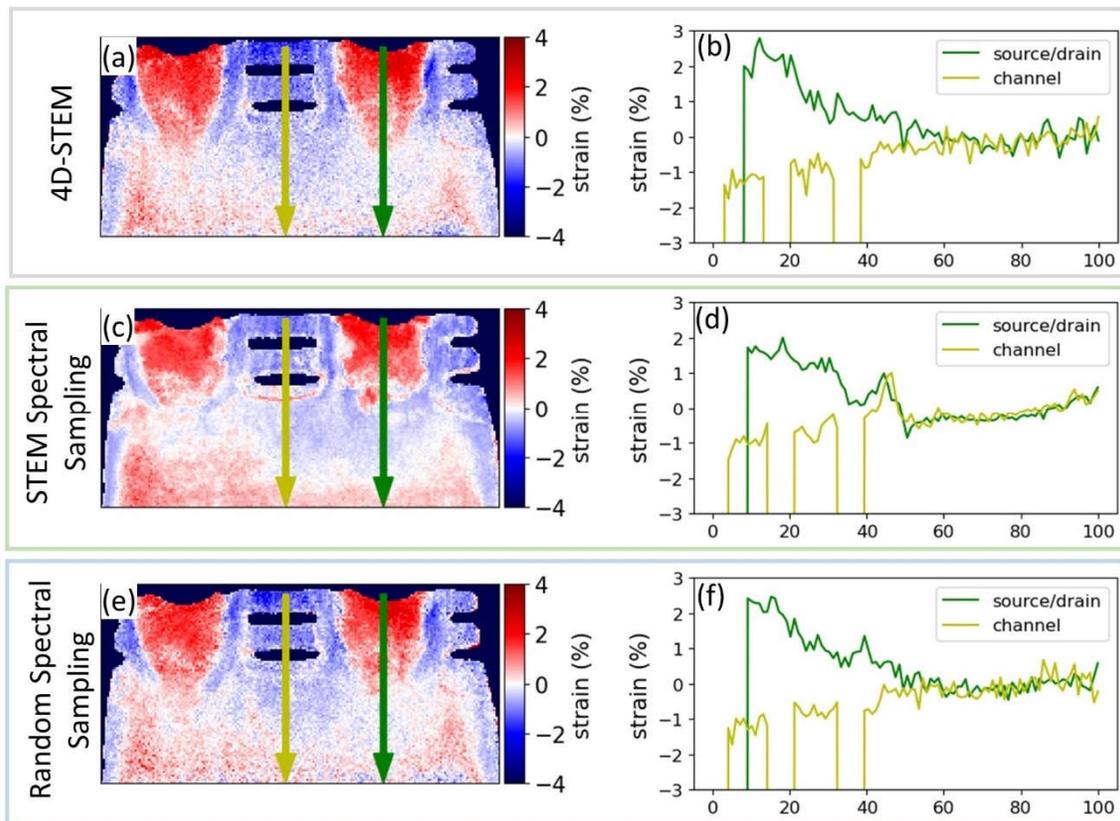

*Figure 4. strain analysis of datasets reconstructed with different spectral sampling method. (a) original dataset (b) STEM spectral sampling with 300 virtual images (c) random spectral sampling with 300 virtual images. Unit for the x-axis in the line-profiles is in nanometer.*

Random spectral sampling can also be used for the compression of 4D-STEM datasets. For the example using 300 virtual images sampling formed by randomly selected detector pixels, the reconstructed 4D-STEM dataset was decomposed into two matrices of 20000 by 83 and 83 by 15376 in dimensions, giving rise to a 100-fold smaller data volume. To implement such scheme, new detector technology with fast random readout is required to provide the capability of randomly sampling in diffraction space with scanning speed close to conventional STEM imaging.

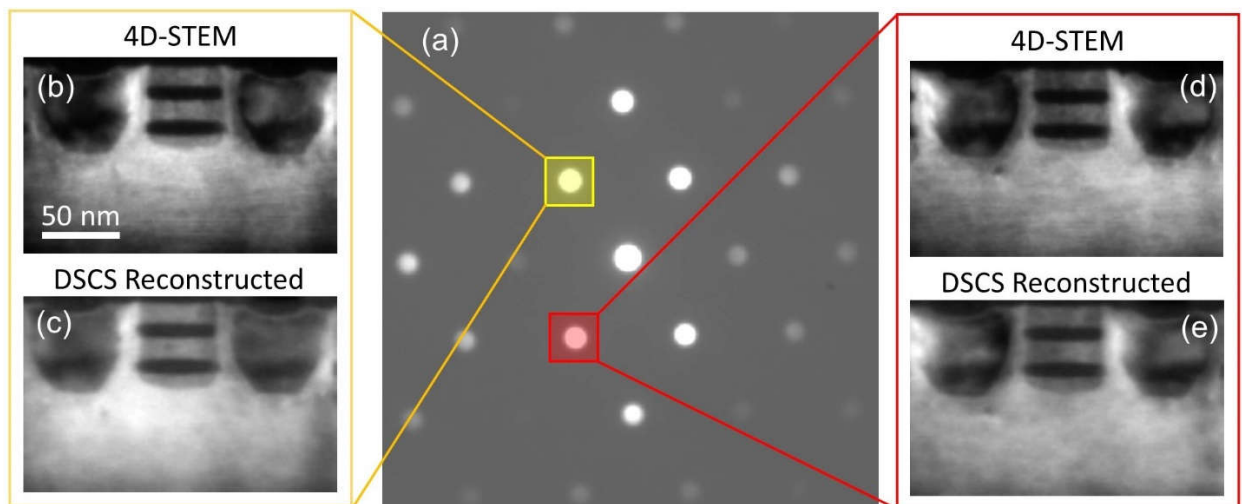

*Figure 5*. DSCS applied for data acquisition in A STEM and comparison with the full 4D-STEM data acquired using EMPAD. *The DSCS dataset was acquired using random scan and STEM images formed using segmented detectors. (a) shows an example DP in the reconstructed 4D-STEM dataset using DSCS. Virtual dark-field images created with the yellow (top) and red (bottom) virtual apertures from (b,c) full 4D-STEM dataset and (c,e) DSCS reconstructed dataset.*

### 3.6 DSCS Implementation in a STEM

To explore experimental feasibility of DSCS, we implemented it on a Themis Z S/TEM (Thermo Fisher Scientific, USA, probe corrected and operated at 300kV) equipped with a Gatan Image Filter (GIF) and an EMPAD detector at University of Illinois. The microscope was operated in STEM mode and the diffraction patterns were collected with the CCD mounted behind the GIF. First, a region of interest (ROI) is selected for imaging. The ROI is then scanned using the Bernoulli sampling discussed in section 2 by modifying the scanning electron nanodiffraction setup described by Kim et al.[42]. At each sampling point, the diffraction pattern and the sampling coordinates are recorded and saved. The control of beam movement and synchronization of camera and scanning was achieved by DigitalMicrograph scripting. A dataset with 5% random spatial sampling was acquired with the custom script. Diffraction space sampling was done by taking STEM images using segmented detectors with 2 different camera lengths to sample different parts of the diffraction pattern. The sampled DPs and STEM images were used to obtain a reconstructed 4D-STEM dataset. At the same ROI, we also acquired a full 4D-STEM dataset using EMPAD for comparison.

Fig. 5 shows a set of virtual dark-field images generated from the reconstructed dataset and the full 4D-STEM dataset. Promisingly, the diffraction contrast is mostly reproduced. The small discrepancies in the images likely due to the experimental factors, such as the difference in the cameras' readout speeds, and the sample drift during DP collection, which makes the spatial registration difficult. This further influence the registration between diffraction patterns and STEM images in DSCS. Additionally, the STEM detectors and CCD cameras have different response to electrons, correlating intensities from these two sources may also introduce noise and error. Lastly, when acquiring STEM images, the position of the diffraction pattern on the plane of STEM detectors was manually aligned so the spectral measurement matrix is estimated rather than precisely determined as in the synthetic CS dataset. These difficulties can be overcome by using the same detector for spatial and spectral sampling.

## 4. Conclusion

We have introduced an approach for 4D-STEM compressive sensing and data compression based on sub-sampling in both real and diffraction spaces. The method exploits the sparsity of diffraction data. We show that a full 4D-STEM dataset can be recreated from partially sampled 4D-STEM data. To demonstrate the efficiency of our method, we created test data by sampling a full experimentally collected 4D-STEM dataset using random scan in the real space and two types of sampling in the reciprocal space. An iterative reconstruction algorithm is developed and used for reconstruction. The performances of the algorithm for the two types of sampling are compared. The results show that a factor of 100 data reduction can be achieved in 4D-STEM data compression with good quality. The quality of DSCS reconstruction is measured by the structural similarity index and peak signal to noise ratio between the original and reconstructed diffraction patterns and the fidelity of reconstructed DPs is tested with strain analysis. The results also show that high-fidelity reconstruction is achieved with only 5% of spatial sampling. These results demonstrate the efficiency of DSCS. For 4D-STEM experiments, the adaptation of compressive sensing will reduce the number of scanning points significantly so that the electron dose is greatly reduced. Applications previously hindered by the high dose and long acquisition time, such as the analysis of beam sensitive materials will become possible with the presented method.

Acknowledgement: This work is supported by a grant from Intel. JMZ is supported by by NSF DMR-2226495 from the Metals and Metallic Nanostructures Program (MMN) within the Division of Materials Research.